\documentclass[aip,reprint, jmp, amsmath, amssymb]{revtex4-1}
\usepackage{graphicx}% Include figure files 
\usepackage{dcolumn}% Align table columns on decimal point
\usepackage{bm}% bold math

\begin{document}

\title{Long coherence time of spin qubits in $^{12}$C enriched polycrystalline CVD diamond} %Title of paper

\author{K. D. Jahnke}
%\email{kay.jahnke@uni-ulm.de.}
% \homepage{http://www.Second.institution.edu/~Charlie.Author.}
\affiliation{Institute for Quantum Optics, Ulm University, Ulm 89081, Germany.}
\author{B. Naydenov}
\email{boris.naydenov@uni-ulm.de.}
\affiliation{Institute for Quantum Optics, Ulm University, Ulm 89081, Germany.}
\author{T. Teraji}%
\affiliation{National Institute for Materials Science, Tsukuba, Ibaraki, 305-0044, Japan.}
\author{S.Koizumi}%
\affiliation{National Institute for Materials Science, Tsukuba, Ibaraki, 305-0044, Japan.}
\author{T. Umeda}%
\affiliation{Research Center for Knowledge Communities, University of Tsukuba, Tsukuba, Ibaraki, 305-8550, Japan.}
\author{J. Isoya}%
\affiliation{Research Center for Knowledge Communities, University of Tsukuba, Tsukuba, Ibaraki, 305-8550, Japan.}
\author{F. Jelezko}
\affiliation{Institute for Quantum Optics, Ulm University, Ulm 89081, Germany.}
\date{\today}

\begin{abstract}
Single defects in diamond and especially negatively charged nitrogen vacancy (NV) centers are very promising quantum systems with wide applications in physics and biology. It was shown that their coherence properties can be strongly improved by growing ultrapure diamond with low concentration of parasitic spins associated with nitrogen electron spins and nuclear spins related to $^{13}$C carbon isotope. Here we report a high quality $^{12}$C-enriched polycrystalline CVD diamond material with properties comparable with single crystals. We find single NVs in the grains of this material, which show extremely long electron spin coherence time $T_2 > 2\,ms$.
\end{abstract}


\maketitle 

\section{Introduction}

During the last decade diamond has attracted attention as one of the promising materials for novel quantum technology applications\cite{Weber10}. It was shown that single optically active defects (color centers) can be detected and controlled using combination of optical microscopy and magnetic resonance techniques\cite{Wrachtrup06}. Negatively charged nitrogen vacancy centers (NVs) in diamond have been intensively studied, since they have an application as quantum bits (qubits)\cite{Neumann10,Jelezko04a,Jelezko04b,Dutt07} and for ultra-sensitive detection of electric\cite{Dolde11} and magnetic fields\cite{Laraoui10,Gopi08,Maze08} with nanometer spatial resolution. These defect centers in diamond consist of a substitutional nitrogen atom next to a vacancy at the adjacent lattice site. This system has a triplet ground state with a strong optical transition towards excited triplet states and shows dependence of the fluorescence emission on spin state, allowing the detection of spin states associated with single defects\cite{Gruber97}. 
The spin coherence time $T_2$ associated with the ground state of NV defects depends on the concentration of paramagnetic centers in the diamond lattice. It was shown that isolated neutral substitutional nitrogen denoted the P1 center\cite{Smith59} in the field of electron paramagnetic resonance(EPR) plays the major role and limits the coherence time of NV defects to a few microseconds in nitrogen-rich type-Ib diamond grown by high pressure high temperature (HPHT) method\cite{Kennedy02}. Recently there has been great improvement in the quality of diamond crystals grown using chemical vapor deposition (CVD) method where the nitrogen concentration has been reduced below $0.1 \,$ppb \cite{Markham11,Balmer09} levels. In such an electron spin free environment coherence time ($0.65\,$ms) is limited only by the magnetic fluctuation related to the $^{13}$C (I=1/2, natural abundance 1.1\%) nuclear spin bath\cite{Mizuochi09}. $T_2$ can be improved even further if the nuclear spin possessing isotope $^{13}$C is depleted and $T_2 = 1.8\, $ms has been already demonstrated for ultra-high pure single crystal (SC) CVD with $^{13}$C = 0.3\%\cite{Gopi09} .
As it was mentioned above spectacular coherence properties were first observed for homoepitaxially grown SC CVD diamond. The size of this crystals is limited by availability of HPHT substrate. More recently it was shown that $^{13}$C spin bath limited coherence time can also be achieved for polycrystalline diamond material\cite{Markham11}. Here we show that isotopically enriched polycrystalline diamond material can achieve record $T_2$ comparable with that of SC CVD diamond.
Polycrystalline CVD diamond films, which are typically obtained when non-diamond substrates are used, can be grown over large areas, thus, obtained in large quantities with lower cost. This material is inhomogeneous, being consisted of grains (small single crystals) and grain boundaries. However, some physical properties of polycrystalline materials are dominantly determined by the grains and not by the grain boundaries. Thus, large-area polycrystalline diamond is technologically important for applications using the unique physical properties of diamond\cite{Balmer09}. Each specific application requires a growth condition which optimizes a given property, such as hardness, transparency, thermal and electronic conductance \cite{Balmer09}.
The physical properties of the grains in polycrystalline diamond are expected to be strongly affected by impurities and defects, as in the case of single crystal. Since the coherence time of the NV center is sensitive to the presence of paramagnetic impurities, defects and the isotope enrichment in the lattice, native-NV, which is a grown-in defect in CVD diamond, is a useful probe for characterizing the quality of the grains. In this study, NIMS home-made microwave plasma-activated chemical vapor deposition (MPCVD) system, that is well designed to minimize external leakage, was used. This system equipped the residual gas evacuation lines, that help to suppress the gas mixture of $^{12}$C enriched methane with commonly used natural abundance methane. In addition, $^{12}$C enriched methane containing 3 ppm nitrogen was purified
by the methane gas filter that reduces the residual impurity level being less than 10 ppb. Secondary ion mass spectrometry (SIMS) measurements proved that $^{12}$C enrichment of the polycrystalline diamond film is 99.998\%\cite{Teraji12}. % In the present work, the samples have been grown using $^{12}$C 99.999\% enriched methane by a microwave plasma-activated chemical vapor deposition (MPCVD) method\cite{Teraji12}. 
Free-standing film (thickness 230 $\mu m$) removed from the molybdenum substrate was used. The nucleation surface is relatively smooth and the growth surface is rough reflecting the evolution of the surface morphology as the film growth.
\begin{figure}
 	\includegraphics[scale=0.3]{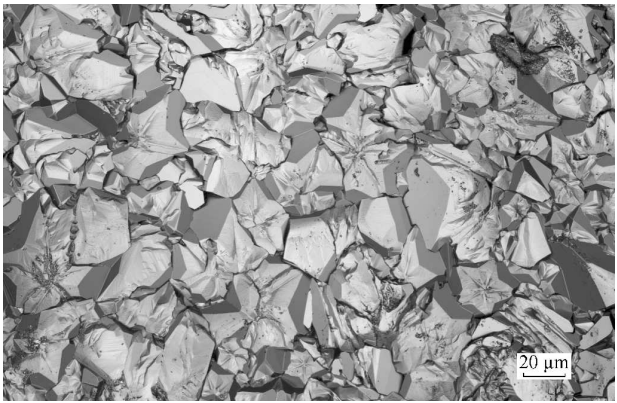}%
 	\caption{\label{SEM} Laser microscope picture of the polycrystalline diamond. Grains with a size of around $30 \,\mu m$ are easily distinguishable.}
\end{figure}	
The size of the grains in this material is around $30 \,\mu m$, as estimated from laser microscope (Keyence Ltd.: VK-9700) measurements. A typical picture is shown in Figure \ref{SEM}. In polycrystalline CVD diamond, preferential orientation of grains is often observed \cite{Balmer09,Kobashi05}. At the initial stage of the growth, the grains are randomly oriented when the density of nucleation sites on the substrate is high. However, the grains aligned along favored orientations grow faster than the rest and fill up the volume first. As a result, the grains which reach a size of a few tens of $\mu m$ at the growth surface, exhibit preferential orientation. In thicker films, columnar growth occurs. This preferred orientation depends on the substrate and the growth conditions such as the CH$_4/$H$_2$ ratio, the addition of nitrogen to the gas mixture, and the substrate temperature. Highly oriented [100]-textured films were obtained in a polycrystalline CVD diamond grown on Si (100) \cite{Kobashi05,Graeff97}. In our samples, the grains have their crystal lattice preferentially aligned with the [110] axis perpendicular to the substrate surface\cite{Teraji12}. In the case of [110] preferential orientation, the in-plane orientations of the grains are random \cite{Kobashi05}. The facets appearing on the surface are not necessarily (110) parallel to the plane of the film and can be (111) which is tilted by $35\,^{\circ}$ from the growth direction\cite{Kobashi05}. The details of synthesis of $^{12}$C enriched polycrystalline CVD diamond films and X-ray diffraction (XRD) studies will be described elsewhere \cite{Teraji12}.
\begin{figure}
	\includegraphics[scale=0.46]{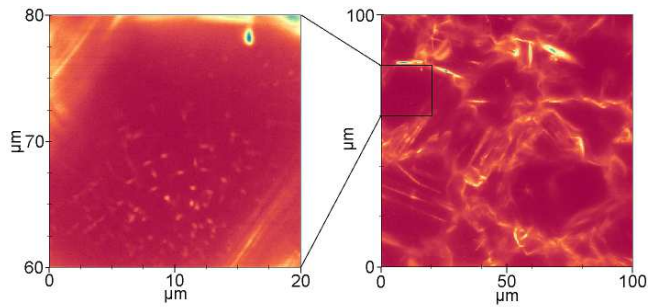}%
	\caption{\label{confocal} Scanned confocal fluorescence image of a polycrystalline CVD Diamond (right) and a zoom into a $20 \,\mu m \times 20 \,\mu m$ region (left). Red and blue color indicate low and high fluorescence respectively. In the left part some single NVs are visible.}
\end{figure}
Isotopically engineered diamond material was characterized using optical microscopy techniques. The fluorescence of the NV centers has been measured on a custom built confocal microscope with an oil immersion objective (NA = 1.4) and excitation laser with a wavelength of $\lambda = 532\, nm$. Scattered laser light was blocked by a 640 nm long-pass filter. Figure \ref{confocal} shows a typical fluorescence image of the growth side of a polycrystalline diamond sample. As it is often observed in this type of material, grain boundaries show strong fluorescence signal owing to increased incorporation of impurities and defects in these regions\cite{Markham11}. They also show disordered structure containing carbon dangling bonds and hydrogen impurities. The size of the individual grains is in a good agreement with the electron microscopy data. Regions far from boundaries show much lower fluorescence background with distinct fluorescence spots related to single NV defects. In order to investigate their optical and spin properties, fluorescence correlation spectroscopy and optically detected magnetic resonance (ODMR) techniques were used. Two avalanche photo diodes (APD) in a Hanbury Brown and Twiss configuration were used as detectors, allowing us to measure the second order fluorescence intensity correlation function g$^{(2)}(\tau)$, thus estimating the number of emitters in a confocal spot. These measurements confirm (data not shown) that individual fluorescent spots indeed correspond to emission from individual NV defects.

Polycrystalline sample with similar quality and NV concentration has been previously reported\cite{Markham11}. In this earlier work the carbon material had the natural abundance of $^{12}$C, which limits the electron spin coherence time to $T_2 = 0.6\, $ms. Isotopical engineering of the diamond allows further improvement of the $T_2$. Coherence properties of single NV centers were measured using pulsed ODMR. The NV electron spin system can be described by the following Hamiltonian:
\begin{equation*}
	H=g\mu_BB_0S_z+D\!\!\left(\!S_z^2-\frac{1}{3}S(S+1)\!\right)+E\!\left(S_x^2-S_y^2\right)+SAI
	\label{hamiltonian1}
\end{equation*}
With the Bohr magneton $\mu_B$, g-factor of the electron spin $g \approx 2.002$, the components of the NV electron spin ($S = 1$) $S_i$ for $i = x, y, z$, the applied constant magnetic field $B_0$, the axial zero field splitting (ZFS) $D = 2876 \,$MHz, the non-axial ZFS due to strain $E$ and the hyperfine interaction to the nitrogen nuclear spin ($I = 1$) $A = 2.5 \,$MHz. Owing to the fast intersystem crossing from the excited state magnetic sublevels with magnetic quantum number $m_S=\pm1$, a high degree of spin polarization in the ground state can be achieved after a few microseconds of optical excitation. Since the fluorescence intensity of the $m_S=0$ state is higher, magnetic resonance can be detected by measuring the photon emission rate when sweeping the microwave field \cite{Jelezko06}. 
\begin{figure}
	\includegraphics[scale=0.4]{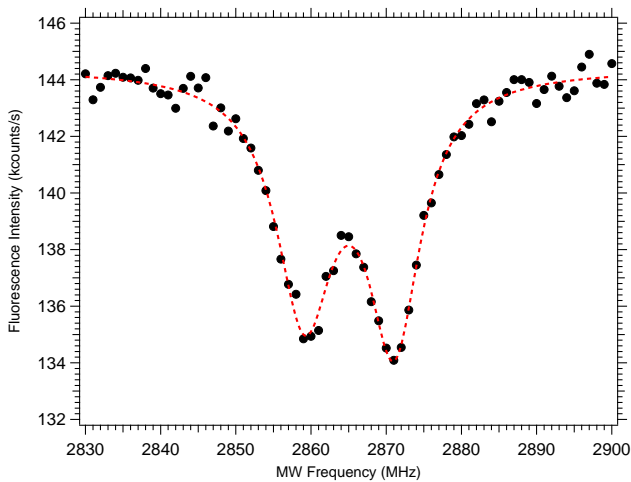}
 	\caption{\label{ODMR}ODMR spectrum of a single NV center. Due to non-axial strain $E$ the degeneracy of the $m_S=\pm 1$ levels is lifted and the line is split.}
 \end{figure}
A typical ODMR spectrum of a single NV center in polycrystalline diamond is plotted in Figure \ref{ODMR}. The spectral line is a doublet centered at the ZFS frequency $\nu_{ZFS} = D$. The spectrum was measured in zero magnetic field. The two peaks correspond to the $m_S =+1 \leftrightarrow m_S = 0$ and $m_S = -1 \leftrightarrow m_S = 0$ transitions which are not degenerated due to non-zero non-axial strain caused by disorder present in the diamond lattice. The red dashed line is a simulation with $E = 12 \, $MHz. All the NVs we analyzed showed significant strain, with an average splitting of $E_{avg} = 20(7)\, $MHz (based on the statistics of 4 centers). Despite of the observed strain, the coherence time of the single NVs is remarkably long for this type of the material. 
\begin{figure}
 	\includegraphics[scale=0.4]{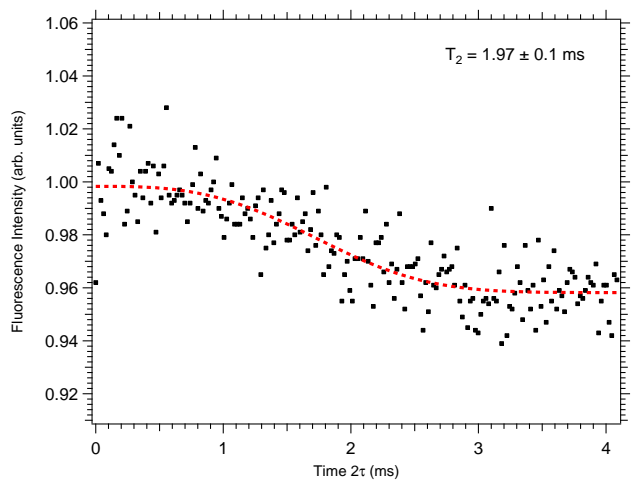}
 	\caption{\label{Hahn} Hahn echo decay of an NV center in polycrystalline diamond (markers). The solid line is a stretched exponential fit to $E(\tau)=a\cdot \exp{\left[-\left(\tau/T_2\right)^2\right]}+b$.}%
\end{figure}
A typical Hahn echo decay from a single NV inside the grains is plotted in Figure \ref{Hahn}. All measured $T_2$ values for different NVs are longer than a millisecond, with an average $T_2 = 1.6 \, $ms. The longest $T_2$ we measured was $T_2 = 2.1 \, $ms, which is slightly longer than the recently reported coherence time of single defects in $^{12}$C enriched (99.7 \%) enriched monocrystalline diamond material\cite{Gopi09}. This result is a bit surprising since a higher concentration of impurities in the polycrystalline CVD diamond is expected compared to the mono-crystal. The relaxation time $T_1$ of the NV is in average 5.3 ms (data not shown) and is also comparable with NVs in single crystal diamond.
The $T_2$ longer than 0.65 ms which is the best one obtained so far in natural abundance samples, requires $^{12}$C enrichment. Ensemble measurements of SC CVD diamond with $^{12}$C natural abundance show $T_2 > 0.6\,ms$ which is limited by the $^{13}$C nuclear spin bath, where a low concentration of the nitrogen substitutional (electron spin S = 1/2) P1-centers of $\approx 10^{15} cm^{-3}$ (10 ppb corresponds to $1.76\cdot 10^{15} cm^{-3}$) is necessary \cite{Stanwix10}. Prolonging $T_2$ by $^{12}$C enrichment requires low total nitrogen concentration in the lattice, while $^{12}$C enriched methane has a substantially higher amount ($>3\,$ppm) of nitrogen impurities compared to methane with natural abundance. The long $T_2$ of $ \approx 2\,$ms in our $^{12}$C enriched polycrystalline CVD diamond suggests that the in-grain concentration of P1 is significantly lowered. 

The concentration of these paramagnetic centers inside the grains was measured by continuous wave (CW) electron paramagnetic resonance (EPR) on a Bruker E500 spectrometer operating at the microwave frequency of 9.65 GHz. Several films (with total weight 60.8 mg) glued in a stack on both sides of high-purity silica glass plate were used as a sample. It is expected that the P1 concentration should be different between the gains and their boundaries and between the randomly-oriented grains at the initial growth stage and those grown up to a preferential orientation. Instead of obtaining the average concentration of P1 from the whole sample, it is necessary to extract its signal selectively from the preferentially oriented grains. It is likely that the P1 center in these grains, where its concentration is low, has longer $T_2$ compared to $T_2$ in other parts of the sample. Moreover, it is expected that the EPR spectrum from there should show a hint for the crystal orientation. EPR spectroscopy of the P1 centers was previously used to characterize preferential orientation of the grains of polycrystalline CVD diamond films\cite{Graeff97,Nokhrin01}.
Usually a slow passage technique is applied for the quantitative measurements, but it lacks sensitivity if the relaxation and coherence time is long. In this case a rapid passage method is used for determining low concentrations (0.1 ppb to 1 ppm) of P1 centers \cite{Cann09}

We have used this technique to measure the amount of P1 of preferentially oriented grains. This center has $C_{3v}$ symmetry, where the unpaired electron occupies the anti-bonding orbital of one of the four N-C bonds. The anti-bonding character is lowered by the elongation of the N-C bond length. There are four symmetry-related sites corresponding to four C-C directions. The EPR spectrum of P1 consists of three lines arising from hyperfine coupling to $^{14}$N (I=1, natural abundance 99.64\%). The central line is nearly isotropic with a small g-anisotropy ($g_{||} = 2.0024$, $g_\bot =2.0025$) \cite{Zhang94}. The positions of the outer lines are predominantly determined by the anisotropy of the hyperfine splitting which is in the first-order given by,
\begin{equation*}
	A_{eff}(\theta)=\left(A_{||}^2 \cos^2\theta + A_{\bot}^2 \sin^2\theta\right)^{1/2}
	\label{hyperfine}
\end{equation*}
where $A_{||}= 4.093 \,mT$, $A_{\bot}=2.920\,mT$, and $\theta$ is the angle between the symmetry-axis and the external magnetic field. In single crystal at an arbitrary direction of the magnetic field, EPR spectrum of P1 consists of four sets of outer lines corresponding to four sites since $\theta$ is different among them.
\begin{figure}
 	\includegraphics[scale=0.37]{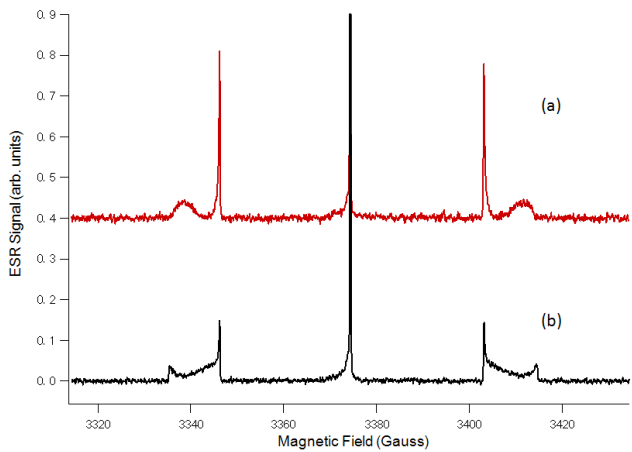}
 	\caption{\label{EPR}Rapid passage cw-EPR spectrum of $^{12}$C enriched polycrystalline CVD film at room temperature. Central line is cut off. (a) magnetic field (\textbf{B}) perpendicular to the film plane, (b) \textbf{B} parallel to the film plane.}
\end{figure}
For [110]-oriented films, when the magnetic field is perpendicular to the film plane, it is expected that the EPR spectrum is similar to one taken for $B||$[110] for single crystal sample in which the hyperfine spectrum consists of two sets, one from two sites with $\theta=90\,^{\circ}$ and the other from two sites with $\theta=35.26\,^{\circ}$. The two sites in each set are magnetically equivalent in this orientation. As shown in Figure \ref{EPR}(a), the EPR spectrum taken with the magnetic field perpendicular to the film plane is similar to that of single crystal spectrum taken with the magnetic field along [110]. In the oriented polycrystalline CVD diamond, there is a distribution of the angles between the crystal axis of each grain and the direction perpendicular to the film plane. 
Thus, the positions of the outer $^{14}$N hyperfine lines ($m_I= \pm1$) are additionally broadened due to the misalignment of the grain axis with respect to the normal to the film plane. Here the grain axis is [110] of the crystal lattice of each grain. The misalignment more strongly affects the line positions of the sites corresponding to $\theta=35.26\,^{\circ}$ compared to those with $\theta=90\,^{\circ}$, resulting in broader lines of these sites.
As shown in Figure \ref{EPR}b, where the magnetic field is parallel to the film plane, a line shape corresponding to random orientations of the grains is observed as expected. 
Thus, the EPR spectrum extracted by the rapid passage arises from P1 centers located in the preferentially oriented grains and from the signal intensity their concentration can be estimated. As shown in Figure \ref{confocal}, the volume contribution of the grain boundaries is relatively small and the preferential oriented grains dominate after growing 10\% of the film. By using the whole volume of the sample, the average concentration of P1 of the preferentially oriented grains is determined to be as low as 4 ppb, explaining the long coherence time of a single NV center in these regions. We note here that the distribution of P1 in each grain is not likely to be uniform and the by us detected NV centers lie within a depth of $\approx 10 \,\mu m$ which corresponds to the top layer of the grains. The large quantity of polycrystalline material has made the measurement of the P1 concentration of such a low concentration possible.\\
In a slow passage condition (microwave power $8\,\mu W$), the EPR spectrum reveals the H$_1^{'}$ center with $g=2.0029$ (data not shown). A weak hyperfine interaction with nearby hydrogen is revealed by the $^1$H nuclear-spin flip transitions. This center is likely to be a carbon dangling bond in the grain boundaries where hydrogen is incorporated.

\section{Conclusion}
In summary, we have demonstrated a long coherence time $T_2$ for single NV centers located in the grains of polycrystalline CVD diamond grown from $^{12}$C 99.999\% enriched methane. The measured $T_2$ exceeding 1 ms requires both depleting the $^{13}$C isotope and lowering the concentration of P1 which is the dominant form of nitrogen impurity in CVD diamond. Since the $^{12}$C enriched methane contains a substantial amount of nitrogen impurity ($> 3 $ppm), it is challenging to obtain a low nitrogen concentration in $^{12}$C enriched CVD diamond, especially in polycrystalline material. We observe single NV defects indicating that concentration of nitrogen incorporated during the crystal growth is lower than ppm level and concentration of NV defects is on the order of 0.1 ppb.
The EPR spectrum using rapid passage technique shows P1 located inside the grains which exhibits a characteristic feature of preferential orientation, where the in-grain concentration is estimated to be 4 ppb.
The coherence time of the single NV defects is comparable with reported $T_2$ of NVs in a SC CVD diamond. Thus, high quality (i.e. low concentrations of paramagnetic impurities and defects) of the grains has been proved. This material can be produced in large quantities with lower costs, compared to SC CVD diamond. This opens many new possibilities, for example the production of large amounts of nanodiamonds containing NVs with good properties by milling the diamond grains. 

This work was partly supported by the DFG (projects: FOR1482, FOR1493, SFB/TR21), the Strategic International Collaborative Research Program (Nanoelectronics) from Japan Science and Technology Agency, EU via project DIAMANT and the Grant-in-Aid for Scientific Research from the Japan Society for the Promotion of Science, Japan (No. 18760241).

\end{document}